\documentclass[preprint,12pt]{article}
\usepackage{amssymb}
\usepackage{graphicx}
\usepackage{multirow}
\usepackage{caption}
\usepackage{subfig}
\begin{document}
\title
{New mass limit of white dwarfs}
\author{Upasana Das and Banibrata Mukhopadhyay\\ \\
Department of Physics, Indian Institute of
  Science, Bangalore 560012, India\\
upasana@physics.iisc.ernet.in , bm@physics.iisc.ernet.in}
\maketitle
\begin{center}
Essay received Honorable Mention
in the Gravity Research Foundation \\ 
2013 Awards for Essays on Gravitation
\end{center}

\vskip1.0cm
\begin{abstract}

Is the Chandrasekhar mass limit for white dwarfs (WDs) set in stone? Not anymore --- recent 
observations of over-luminous, peculiar type~Ia supernovae can be explained if 
significantly 
super-Chandrasekhar WDs exist as their progenitors, thus barring them to be used 
as cosmic distance indicators. However, there is no estimate of a
mass limit for these super-Chandrasekhar WD candidates yet. Can they be arbitrarily 
large? In fact, the answer is no! We arrive at this revelation by exploiting the 
flux freezing theorem in observed, accreting, magnetized WDs, which 
brings in Landau quantization 
of the underlying electron degenerate gas. This essay presents the calculations which 
pave the way for the ultimate (significantly super-Chandrasekhar) 
mass limit of WDs, heralding a paradigm shift 
$80$ years after Chandrasekhar's discovery.

\end{abstract} 

\indent \indent equation of state of gases \\

\hskip0.5cm PACS Number(s): 97.20.Rp, 97.60.Bw, 97.10.Ld, 71.70.Di, 51.30.+i \\ \\


\noindent \textbf{Introduction}
\vskip0.5cm

\noindent 
Chandrasekhar, in one of his celebrated papers \cite{chandra}, showed that the maximum
possible mass of non-rotating, non-magnetized white dwarfs (WD) is $1.44M_\odot$, when 
$M_{\odot}$ being the mass of Sun. He was awarded the Nobel Prize in Physics in 1983 
mainly because of this discovery.
This limiting mass (LM) is directly related to the luminosity of observed type~Ia supernovae which 
are used as standard candles for measuring far away distances and hence in understanding 
the expansion history of the universe. The discovery of the accelerated 
expansion of the universe led to the Nobel Prize in Physics in 2011 \cite{pelmutar}.

So far, observations seemed to abide by the Chandrasekhar limit. However, in order to 
explain the recent 
discovery of several peculiar, anomalous, distinctly over-luminous type~Ia supernovae 
\cite{howell,scalzo} -- 
namely, SN~2006gz, SN~2007if, SN~2009dc, SN~2003fg -- the mass of the exploding WDs 
(progenitors of supernovae) needs to be between $2.1-2.8M_\odot$, significantly super-Chandrasekhar.
Nevertheless, these non-standard `super-Chandrasekhar supernovae' can no longer be used 
as cosmic distance indicators. However, there is need of a foundational level analysis, 
akin to that carried out by Chandrasekhar, in order to establish a 
super-Chandrasekhar mass WD. Moreover, there is no estimate of an upper mass limit 
for these super-Chandrasekhar WD candidates yet. Can they be arbitrarily large? 
These are some of the fundamental questions, we plan to resolve in the present essay.

\vskip1.0cm

\noindent \textbf{Basic physical process rendering the new limit}
\vskip0.5cm



\noindent We plan to exploit the effects of magnetic field to establish the new limit.
Hence, we consider the collapsing star to be magnetized and 
the resulting accreting WD to be highly magnetized. 
This is in accordance with observations, which show that about $25\%$ of accreting WDs, 
namely cataclysmic variables, are found to have surface magnetic fields as high as 
$10^7-10^8$ G \cite{wick}. Hence their expected central fields could be $2-3$ 
orders of magnitude higher. 
If a magnetized WD gains mass due to accretion, 
its total mass increases which in turn increases the gravitational power and hence 
the WD contracts in size. However, the total magnetic flux in a WD,
$\propto  B R^2$, when $B$ is the magnetic field and $R$ the WD's radius, is conserved. 
Therefore, if the WD shrinks, its radius decreases and hence magnetic field increases. 
This in turn increases the outward force balancing the increased inward gravitational force, 
leading to a quasi-equilibrium situation. As accretion is a continuous process, the
above process continues in a cycle and helps in increasing $B$ above the critical value 
$4.414\times 10^{13}$ G to bring in Landau quantization effects \cite{ud12}. Subsequently, 
the mass of the WD keeps increasing, even above the Chandrasekhar limit,
until the gain of mass becomes so great that it attains a new 
limit. At this point the total outward pressure is unable to support the gravitational 
attraction any longer, leading to a supernova explosion.
This we argue to observe as a peculiar, over-luminous type Ia supernova, 
in contrast to their normal counter parts.

\vskip1.0cm
\noindent \textbf{Computing the new mass limit}
\vskip0.5cm

\noindent In the presence of strong magnetic field, the equation of state (EoS) of degenerate electron gas for
WDs can be recast, at least in the piecewise zones of density ($\rho$), in the polytropic 
form: $P=K_m\rho^\Gamma$,
when $P$ is the pressure, $K_m$ and $\Gamma=1+1/n$ are 
piecewise constants in different regimes of $\rho$ \cite{ud12}. At the highest density regime 
(which also corresponds to the highest magnetic field regime), $\Gamma=2$. Now we 
recall the 
condition for magnetostatic equilibrium and estimate of mass, assuming the WD 
to be spherical, as
\begin{eqnarray}
\frac{1}{\rho}\frac{d}{dr}\left(P+\frac{B^2}{8\pi}\right)=F_{gr}+\left.\frac{\vec{B}\cdot\nabla\vec{B}}
{4\pi\rho}\right|_r,\,\,\,\,\frac{dM}{dr}=4\pi r^2\rho,
\label{magstat}
\end{eqnarray}
when $r$ is the radial distance from the center of WD and $F_{gr}$ the gravitational force.
We note here that the choice of a Newtonian framework is justified in our case as the 
density corresponding to the degenerate pressure is much smaller than the matter density 
to contribute significantly to the effective mass of 
the WD (see Figure 1 of \cite{ud12}). Moreover, in order to correctly include the effect 
of strong magnetic field in a general relativistic hydromagnetic balance equation, 
the Tolman-Oppenheimer-Volkov (TOV) equation, which is true only for the non-magnetic case, 
itself has to be modified first.
Assuming $B$ varies very slowly around the center of the WD which is the regime of interest,
we obtain \cite{arc}
\begin{equation}
\frac{1}{r^{2}}\frac{d}{dr}\left(\frac{r^{2}}{\rho}\frac{dP}{dr} \right) = -4\pi G \rho,
\label{diff}
\end{equation}
where $G$ is Newton's gravitation constant. This further can be recast, with the use of EoS, 
into
\begin{equation}
\frac{1}{\xi^{2}}\frac{d}{d\xi}\left(\xi^{2} \frac{d\theta}{d\xi} \right) = - \theta^{n}\,\,\,\,
{\rm with}\,\,\,\,\rho = \rho_{c}\theta^{n},
\label{lane}
\end{equation}
where $\theta$ is a dimensionless variable
and
\begin{equation}
\xi = r/a,\,\,\,\,
a = \left [\frac{(n+1)K_m\rho_{c}^{\frac{1-n}{n}}}{4\pi G} \right ]^{1/2}.
\label{a}
\end{equation}
Equation (\ref{lane}) can be solved with the boundary conditions
\begin{equation}
\theta(\xi = 0) = 1,\,\,\,\
\left (\frac{d\theta}{d\xi} \right)_{\xi=0} = 0.
\label{bc2}
\end{equation}
Note that for $n < 5$, $\theta$ becomes zero for a finite value of $\xi$, say $\xi_{1}$, 
which basically corresponds to the surface of the WD such that its radius 
\begin{equation}
R = a\xi_{1}.
\label{R}
\end{equation}
Also the mass of the WD can be obtained as
\begin{equation}
M = 4\pi a^{3} \rho_{c}\int \limits_{0}^{\xi_{1}} \xi^{2}\theta^{n}\, d\xi.
\label{M}
\end{equation}
Now, the scalings of mass and radius 
of the WD with its central density ($\rho_c$) are easily obtained as
\begin{eqnarray}
M \propto K_m^{3/2} \rho_c^{(3-n)/2n},\,\,\,\, R\propto K_m^{1/2}\rho_c^{(1-n)/2n}. 
\label{scal}
\end{eqnarray}
Clearly $n=3$ ($\Gamma=4/3$) corresponds to $M$ independent of $\rho_c$ (provided $K_m$ is 
independent of $\rho_c$)
and hence LM. Therefore, we have to find out the condition for which $n=3$ and the corresponding 
proportionality constant for the scaling of $M$. 
Note, however, that $n=1$ for the extremely magnetized, highly dense, degenerate electron gas EoS,
which is the present regime of interest. 
Below we explore the generic mass limit of WDs considering two scenarios. 

\newpage
\noindent \textit{$\bullet$ Modeling WDs with varying magnetic field inside}
\vskip0.5cm

\noindent Magnetized WDs are likely to have a varying $B$ profile, 
with an approximately constant field in the central region (CR), falling off 
from the central to surface region.
As WDs evolve by accreting mass, their central and surface $B$s, along with the density, 
increase. This enables the WDs to hold more mass and hence they
deviate from Chandrasekhar's mass-radius relation. Note that a large $B$ corresponds to a
large outward magnetic pressure (along with a magnetic tension). Hence, an equilibrium solution depends on the 
nature of variation of $B$ within the WD such that it might no longer remain spherical. 
Moreover, how fast the field inside the WD decays to a smaller surface value affects its mass and radius. In 
order to give rise to a stable super-Chandrasekhar WD, the 
field needs to remain constant up to a certain region from the center, so that enough mass is accumulated due to 
the Landau quantized EoS.
However, at the limiting 
(very large) density, when the mass becomes independent of (central) density,
the basic trend of the mass-radius relation has to be same as that of Chandrasekhar,
except for the larger mass, in order to achieve $n=3$ in EoS. At this situation, WDs
will become, theoretically, very small such that $R=0$ (and hence the spherical assumption
of its shape does not alter the result), as that obtained by
Chandrasekhar in the absence of $B$. Hence, WDs close to the LM 
practically should have a constant $B$ throughout.
However, at high density, $K_m\propto B^{-1}\propto\rho_c^{-2/3}$, unlike the non-magnetized 
EoS when $K_m$ is independent of $B$ (and $\rho_c$). Hence, when $\rho\rightarrow\rho_c$, 
the highly magnetized EoS reduces to
\begin{equation}
P=K\rho^{4/3}\,\,\,\,{\rm when}\,\,\,\,K=\frac{c\hbar\pi^{2/3}}{2^{1/3}(m_H\mu_e)^{4/3}},
\label{eosc}
\end{equation}
where $c$ is the speed of light, $\hbar$ the reduced Planck's constant, $m_H$ the mass of proton,
$\mu_e$ the mean molecular weight per electron.
Now combining equations (\ref{a}), (\ref{M}) for $K_m=K$ and $n=3$, we obtain the LM
\begin{equation}
M_{l1}=\frac{10.114}{\mu_e^2}\left(\frac{c\hbar}{G m_H^{4/3}}\right)^{3/2}.
\label{mass2}
\end{equation}
For carbon-oxygen WDs, $\mu_e=2$ and hence the LM becomes $4.67M_\odot$.

\vskip0.5cm
\noindent \textit{$\bullet$ Modeling the central part of WDs}
\vskip0.5cm

\noindent Here we consider only the CR of WDs and estimate the mass of this region. 
Of course the size of the CR changes as WDs evolve \cite{rao}, which in turn
determines how underestimated our result is with respect to the total mass (and total radius) 
of WDs. In CR \cite{ud13}
\begin{equation}
K_m=K\rho_c^{-2/3},
\end{equation}
and hence from equation (\ref{scal})
\begin{equation}
M\propto\rho_c^{3(1-n)/2n},\,\,\,R\propto\rho_c^{(3-5n)/6n},
\label{massf}
\end{equation}
revealing $M$ independent of $\rho_c$ for $n=1$, when
the radius becomes independent of the mass in the mass-radius
relation \cite{rao,ud13}.
Now combining equations
(\ref{a}), (\ref{M}) with $n=1$, we obtain the value of LM
\begin{equation}
M_{l2}=\frac{5.564}{\mu_e^2}\left(\frac{c\hbar}{G m_H^{4/3}}\right)^{3/2}.
\label{mass3}
\end{equation}
For $\mu_e=2$ the LM becomes $2.58M_\odot$. 
Note that $M_{l2}$ is arrived at by considering a constant $B$ and hence is naturally 
smaller than $M_{l1}$ which additionally counts the mass accumulated 
outside the CR, for a varying $B$.

%
%
%
%



\vskip1.0cm
\noindent \textbf{Conclusions}
\vskip0.5cm

\noindent We summarize the findings of this work as follows: 
\begin{itemize}

\item More than $80$ years after the proposal of Chandrasekhar mass limit, this new limit perhaps 
heralds the onset of a paradigm shift. 

\item 
The masses of WDs are measured from their luminosities assuming Chandrasekhar's mass-radius 
relation, as of now. These results may have to be re-examined based on the new mass-radius 
relation, at least for some peculiar objects (e.g. over-luminous type~Ia supernovae).

\item Some peculiar known objects, like magnetars (highly magnetized compact objects, 
supposedly neutron stars, as of now), should be examined based on the above considerations, 
which could actually be super-Chandrasekhar WDs.

\item This new mass limit should lead to establishing the underlying peculiar supernovae 
as a new standard candle for cosmic distance measurement. 

\item In order to correctly interpret the expansion history of the universe (and then dark energy), 
one might need to carefully sample the observed data from the supernovae explosions, 
especially if the peculiar type~Ia supernovae are eventually found to be enormous in number. 
However, it is probably too early to comment whether our discovery has any direct implication 
on the current dark energy scenario, which is based on the observation of ordinary type~Ia supernovae. 

\item Importantly, 
one now needs to carry out complete self-consistent 
calculations for the structure of WDs by generalizing the TOV
equation to account for the strong magnetic field 
and pressure anisotropy, which has not been performed yet, in order to confirm
the LM of WDs derived here.

\end{itemize}


\begin{thebibliography}{10}

\bibitem{chandra}
Chandrasekhar, S., MNRAS \textbf{95,} 207 (1935).

\bibitem{pelmutar}
Perlmutter, S., et al., ApJ \textbf{517,} 565 (1999).

\bibitem{howell}
Howell, D.A., et al., Nature \textbf{443,} 308 (2006).

\bibitem{scalzo}
Scalzo, R.A., et al., ApJ \textbf{713,} 1073 (2010).

\bibitem{wick}
Wickramasinghe, D.T., Ferrario, L., Pub. Ast. Soc. Pac. \textbf{112,} 873 (2000).

\bibitem{ud12}
Das, U., Mukhopadhyay, B., Phys. Rev. D \textbf{86,} 042001 (2012).

\bibitem{arc}
Choudhuri, A.R., Astrophysics for Physicists. Cambridge Univ. Press, New York (2010).


\bibitem{rao}
Das, U., Mukhopadhyay, B., Rao, A.R., ApJL \textbf{767,} 14 (2013).

\bibitem{ud13}
Das, U., Mukhopadhyay, B., Phys. Rev. Lett. \textbf{110,} 071102 (2013).































\end{thebibliography}
\end{document}